\documentclass[aip,jmp,amsmath,amssymb,preprint]{revtex4-1}

\pdfoutput=1

\usepackage{graphicx}
\usepackage{dcolumn}
\usepackage{bm}
\usepackage{epsfig}
\usepackage{inputenc}
\usepackage{color}
\usepackage{array}
\usepackage{pstricks}
\usepackage{amsmath,amssymb,mathrsfs}
\usepackage{pict2e}
\usepackage{supertabular}
\usepackage{soul}
\begin{document}
\preprint{AIP/123-QED}
\title{On the two-dimensionalization of quasi-static MHD turbulence}

\author{B. Favier}
\email[]{benjamin.favier@ncl.ac.uk}
\homepage{http://sites.google.com/site/bfavierhome/}
\author{F.S. Godeferd}%
\author{C. Cambon}
\affiliation{LMFA UMR 5509 CNRS, \'Ecole Centrale de Lyon, Universit\'e de Lyon, France}
\author{A. Delache}
\affiliation{LaMUSE EA 3989, Universit\'e Jean Monnet, Saint-\'Etienne, France}
\date{\today}
%
\begin{abstract}
We analyze the anisotropy of turbulence in an electrically conducting fluid in the presence of a uniform magnetic field, for low magnetic Reynolds number, using the quasi-static approximation. 
In the linear limit, the kinetic energy of velocity components normal to the magnetic field decays faster than the kinetic energy of component along the magnetic field [Moffatt, JFM 28, 1967].
However, numerous numerical studies predict a different behavior, wherein the final state is characterized by dominant horizontal energy.
We investigate the corresponding nonlinear phenomenon using Direct Numerical Simulations. \textcolor{black}{The initial temporal evolution of the decaying flow indicates that the turbulence is very similar to the so-called ``two-and-a-half-dimensional'' flow} [Montgomery \& Turner, Phys. Fluids 25(2), 1982] and \textcolor{black}{we offer an explanation for the dominance of horizontal kinetic energy.}
\end{abstract}

\maketitle
%
%
\section{Introduction}

In most geophysical and astrophysical flows, turbulence is affected by forces
that distort significantly some of its scales in an anisotropic manner, such
as the Lorentz force arising from the presence of an external magnetic field
in a conducting fluid.
This specific turbulent dynamics forced by an imposed magnetic field is found in liquid
metals flows, be they of geophysical nature---the melted iron core of the earth---or
of academic interest in the laboratory \cite{alem79}.
\textcolor{black}{
More recent laboratory experiments use sodium or gallium, whereas liquid sodium is used in industrial configurations (such as a French fast breeder reactor Superph\'enix).}

\textcolor{black}{
Generally, the motion of turbulent liquid metals is governed by
magnetohydrodynamics (MHD): an induction equation for the fluctuating
magnetic field ought to be coupled to the Navier-Stokes equations,
and the latter equations are modified in turn by the Lorentz force,
as a feedback from the magnetic field.
In the presence of an external
magnetic field, such MHD coupling results
in new dissipative terms, of Ohmic nature, and selectively
damped waves, the Alfv\'en waves \cite{moff67}.
In these cases (liquid metal), the magnetic diffusivity in the induction equation is greater than the  molecular
viscosity in the Navier-Stokes equations (\textit{i.e.} the magnetic Prandtl number is small compared to one).}

\textcolor{black}{
As discussed in section II, if, in addition, the magnetic Reynolds number
is small enough, the linear regime no longer admits Alfv\'en waves solutions,
and the effect of the Lorentz force reduces to anisotropic
ohmic (or Joule) dissipation term. In this approximation,
the induction equation is so drastically simplified that it does not require a specific study at all.}

Forgetting the possible inhomogeneities arising from the presence of 
boundaries or interfaces in the latter flows, the fact is that homogeneous anisotropic turbulence 
remains far less studied than isotropic turbulence. 
\textcolor{black}{In the context of low magnetic Reynolds numbers,} the response of initially isotropic turbulence 
to a static magnetic field is nonetheless documented, 
in pioneering theoretical works \cite{moff67}, 
many numerical \cite{schu76,zika98,voro05,bura08} and experimental studies \cite{alem79}. 
One of the main properties of this kind of flow is the suppression of the three-dimensional 
motion due to anisotropic linear Joule dissipation, leading to a flow without variations 
in the direction of the imposed magnetic field. In the linear \textcolor{black}{and inviscid} regime, this final state is 
characterized by the following scaling of the vertical fluctuating
velocity component $w$, along the magnetic field, and the horizontal ones $u$ and $v$:\cite{moff67}
\vspace{-1mm}
\begin{equation}
\label{eq:moff_pred}
\left<w^2\right>\simeq\left<u^2\right>+\left<v^2\right> .
\end{equation}
However, recent numerical simulations of low magnetic Reynolds number turbulence show that
 the horizontal kinetic energy is dominant, at least at large scales \cite{voro05,bura08_2}.
\textcolor{black}{In the latter articles, the departure from equation \eqref{eq:moff_pred} may be due to forcing schemes used to maintain the turbulence in a quasi-steady state. Another explanation proposed by Knaepen \textit{et al.} \cite{knae04} is that molecular viscosity might play a role in the decay of quasi-static MHD turbulence. Note that the decrease of the kinetic energy along the magnetic field has also been reported in a model based on the quasi-static
approximation, but incorporating a non-isotropic model for viscous dissipation \cite{kass99}.}
In this article, we present results of Direct Numerical Simulations (DNS) in order to analyze
this nonlinear phenomenon, 
which is up to now considered as a restoration of isotropy,  
but still has to be elucidated \cite{knae08}.

\section{Governing equations and numerical simulations}

We consider initially isotropic homogeneous turbulence in an incompressible conducting fluid, in which $u \simeq v \simeq w$
The fluid is characterized by a kinematic viscosity $\nu$, a density $\rho$ and a 
magnetic diffusivity $\eta=(\sigma\mu_0)^{-1}$; $\sigma$ is the electrical conductivity, 
$\mu_0$ the magnetic permeability. The initial \textit{rms} velocity is $u_0$, the integral length scale $l_0$.
The Reynolds number and its magnetic counterpart are 
 $Re=(u_0l_0)/\nu\gg1$ and $R_M=(u_0l_0)/\eta\ll1$. 
The flow is submitted to a uniform vertical magnetic field $\bm{B}_0$ scaled as Alfv\'en speed as
 $\bm{B}_0=\bm{B}/\sqrt{\rho\mu_0}$. The ratio between the eddy turnover time and the Joule time 
is the magnetic interaction number $N=(B_0^2l_0)/(\eta u_0)$.
Within the quasi-static approximation, which implies that the asymptotic limit $R_M$ goes to zero, but is nonetheless valid for $R_M<1$,\cite{knae04} the Navier-Stokes equations become
\begin{equation}
\label{eq:momentum}
\frac{\partial\bm{u}}{\partial t}+\bm{u}\cdot\nabla\bm{u}=-\frac{1}{\rho}\nabla p+\nu\nabla^2\bm{u}+\underbrace{\frac{B_0^2}{\eta}\Delta^{-1}\frac{\partial^2\bm{u}}{\partial z^2}}_{\bm{F}}
\end{equation}
where $\bm{F}$ is the rotational part of the Lorentz force, $\Delta^{-1}$ is the inverse of the Laplacian operator and $z$ the vertical coordinate, along the direction of $\bm{B}_0$. Compressible effects are not taken into account here, so that $\nabla\cdot\bm{u}=0$.
\begin{figure}[t!]
\includegraphics[width=\columnwidth]{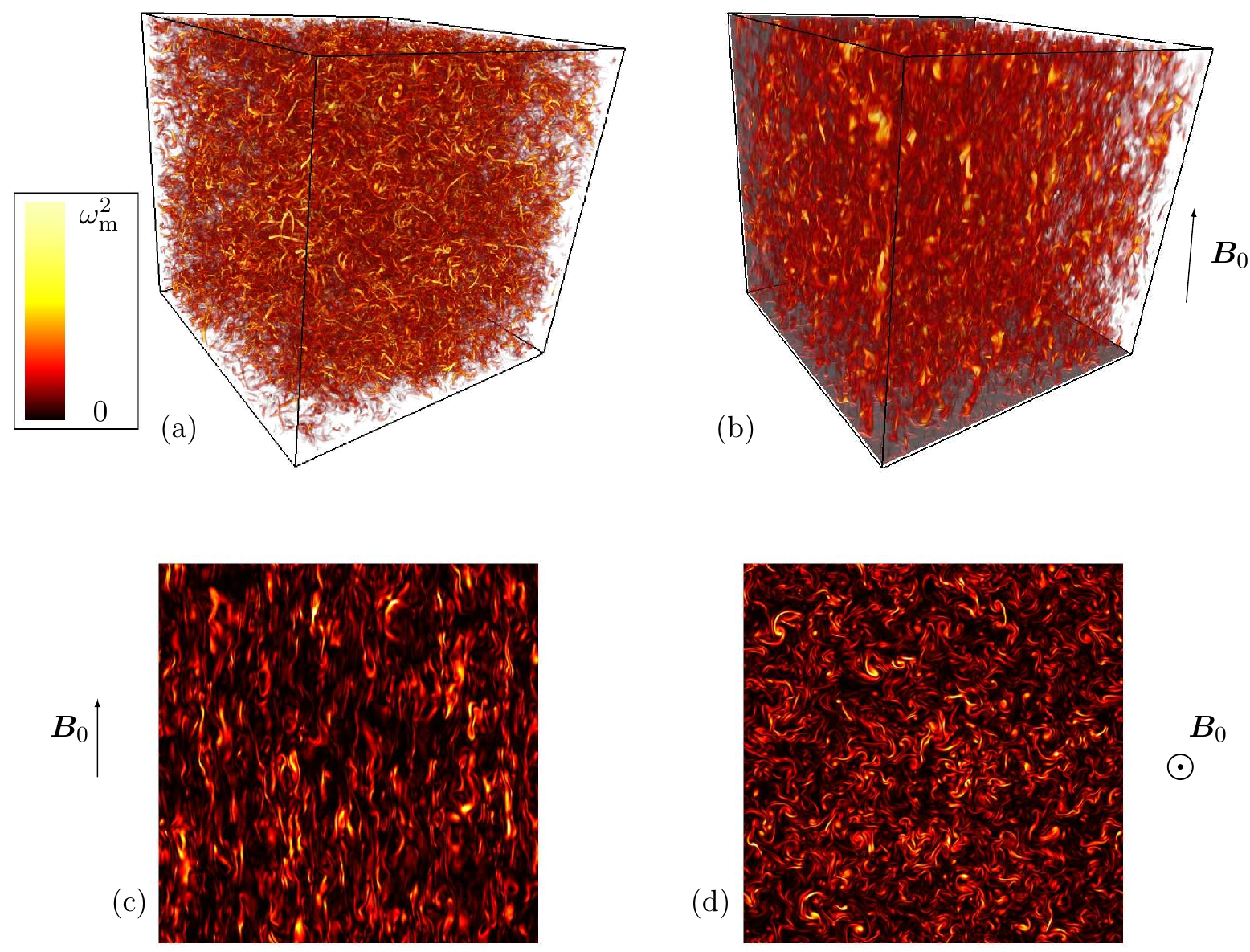}
\caption{\label{fig:visus} Volume rendering of the enstrophy ($\omega_{\textrm{m}}^2$ is about 25\% of the maximum value) using VAPOR \cite{vapor1}. The resolution is $512^3$ (see section \ref{sec:2d3c}). (a) Initial isotropic condition. (b) Final state for $N=5$ (the vertical correlation length is roughly one tenth of the numerical box). (c) Vertical plane extracted from figure (b). (d) Horizontal plane extracted from figure (b).}
\end{figure}
A pseudo-spectral method is used to solve equation \eqref{eq:momentum}. 
The velocity field is computed in a cubic box of side $2\pi$ with periodic boundary conditions 
using $256^3$ Fourier modes. A spherical $2/3$-truncation of Fourier modes is used 
to avoid aliasing and the time scheme is third-order Adams-Bashforth. The dissipative terms are solved implicitly.
The Eulerian velocity field is initialized using an isotropic pre-computation of eq.\eqref{eq:momentum}
 in the hydrodynamic case, \textit{i.e.} $B_0=0$.
The initial incompressible velocity field is a 
random superposition of Fourier modes  distributed with a narrow-band kinetic energy spectrum 
$E(k,t=0)\simeq k^4\exp(-2(k/k_i)^2)$ peaked at $k_i$. Due to the imposed magnetic field, the vertical velocity correlation lengths quickly increase. We therefore avoid interferences with the periodic boundary conditions by choosing $k_i$ larger 
than for a hydrodynamic simulation.
At the small-scale end of the spectrum, the minimum value of $k_{\textrm{max}}l_{\eta}$ is $1.2$ for all our computations \cite{jime93}, with $l_{\eta}$ the Kolmogorov length scale.
At the end of the pre-computation stage, the \textit{rms} velocity is $u_0=0.45$ and 
the integral scale  $l_0=0.16$ yielding $\textit{Re}\simeq100$. \textcolor{black}{In the context of quasi-static MHD turbulence, it is not possible, at least using DNS, to perform high Reynolds number simulations without artificial effect of the periodic boundary conditions. We adopt here an intermediate configuration with moderate value of the Reynolds number and with no noticeable effect of the boundary conditions (see discussion on section \ref{sec:corr_l}). Note that spectral closures such as EDQNM may propose an interesting alternative, reaching high Reynolds numbers and forgetting about possible confinement effects from boundary conditions\cite{camb93}.}

The corresponding turbulent flow field is used as initial state for three different MHD simulations. In all of them $R_M\simeq0.1$ (hence $\eta\simeq1$), so that the quasi-static approximation is justified\cite{knae04}.
Three different amplitudes of the imposed magnetic field are chosen, that correspond to three values of the interaction parameter:
 $N=1$, $3$ and $5$. For each case, 
we perform additional ``linear'' simulations by neglecting the nonlinear advective 
term in eq.\eqref{eq:momentum}. The simulations are freely decaying, since forcing would hinder the natural development of anisotropy, which is the focus of this work.
\begin{figure}
\includegraphics[width=\columnwidth]{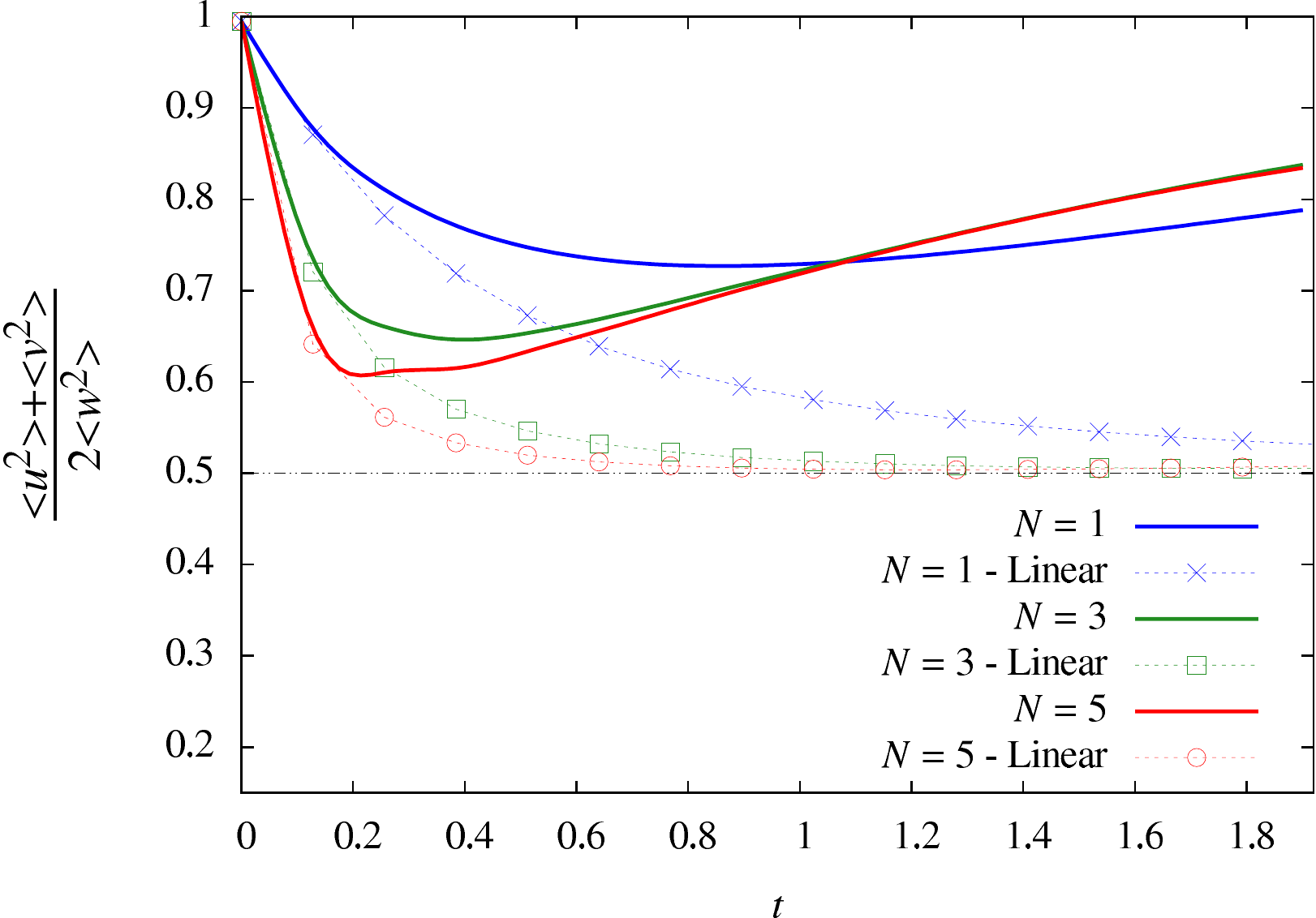}
\caption{Ratio between horizontal (perpendicular to the imposed magnetic field) and vertical (along $\bm{B}_0$) kinetic energy versus time. Continuous lines correspond to nonlinear simulations whereas dotted lines correspond to linear simulations (\textit{i.e.} the non-linear advective term in eq.\eqref{eq:momentum} is removed).}
\label{fig:uvw}
\end{figure}

\section{Results for the Quasi-static case}

\subsection{Kinetic energy and Reynolds stress tensor}

\textcolor{black}{Figure \ref{fig:uvw} presents the ratio between vertical and horizontal energy for different interaction parameters. The linear state characterized by eq.~\eqref{eq:moff_pred} is indeed observed when nonlinear interactions are removed. In the nonlinear simulations, the horizontal kinetic energy decays slower than the vertical one, which is consistent with previous observations \cite{voro05,bura08_2}.}

\begin{figure}
\includegraphics[width=\columnwidth]{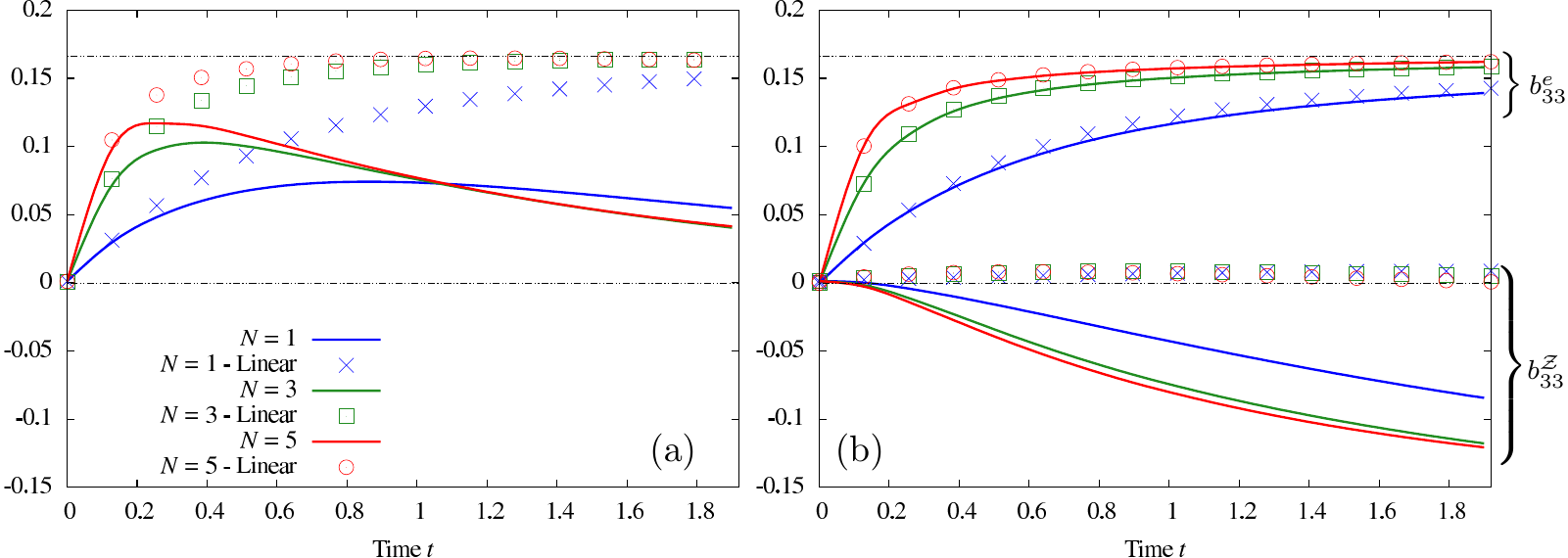}
\caption{Anisotropy tensor of the Reynolds stresses versus time. (a) $b_{33}(t)$. (b) $b_{33}^e(t)$ and $b_{33}^{\mathcal{Z}}(t)$.}
\label{fig:b33}
\end{figure}

\begin{figure}[t!]
\includegraphics[width=\columnwidth]{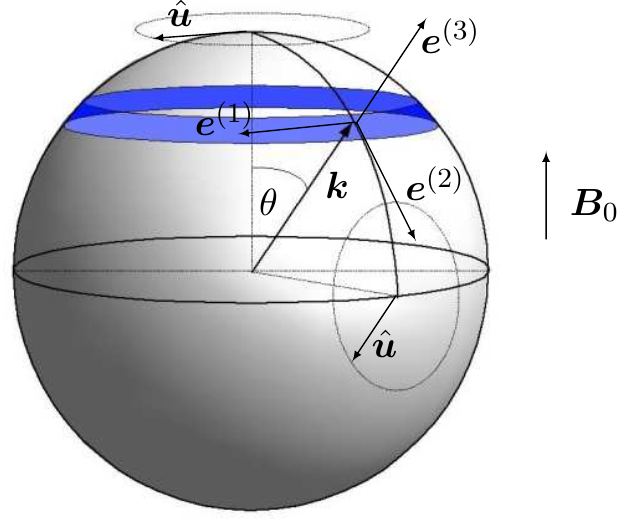}
\caption{Craya-Herring frame $(\bm{e}^{(1)},\bm{e}^{(2)},\bm{e}^{(3)})$ in Fourier space.  Fourier modes 
in the blue region contribute to  $E(k,\theta)$ (eq.\eqref{eq:ekt_def2}). The polar modes (\textit{i.e.} $\theta\simeq 0$) contribute to horizontal kinetic energy, whereas equatorial modes (\textit{i.e.} $\theta\simeq\pi/2$) contribute to both vertical (along $\bm{e}^{(2)}$) and horizontal (along $\bm{e}^{(1)}$) \textcolor{black}{kinetic} energies.
}%
\label{fig:scheme_ezh}
\end{figure}

\textcolor{black}{In order to identify the origin of this phenomenon,} we study the evolution of the Reynolds stress tensor $R_{ij}=\left<u_i(\bm{x})u_j(\bm{x})\right>$ 
and its anisotropic contents $b_{ij}=R_{ij}/(2\mathcal{K})-\delta_{ij}/3$ 
where $\mathcal{K}$ is the total kinetic energy and $\delta_{ij}$ is the Kronecker symbol.
Considering the axisymmetry of the flow about the axis of $\bm{B}_0$, 
only one of the diagonal terms is relevant\footnote{In axisymmetric cases, only the $b_{33}$ component is useful since all the others components are linked by the following relation \cite{CAMBON-MANSOUR-GODEFERD}: $b_{ij}^{e,\mathcal{Z}}=-3(\delta_{ij}/3-\delta_{i3}\delta_{j3})b_{33}^{e,\mathcal{Z}}/2$.}.
Figure \ref{fig:b33}(a) plots the time-dependent $b_{33}(t)$ for the three values of $N$. The index $3$ stands for the vertical direction. The initial value $b_{33}(t=0)=0$ is characteristic of isotropic turbulence. 
In the linear regime (symbols in figure \ref{fig:b33}(a)), 
  $b_{33}$ quickly recovers $b_{33}\simeq1/6$ corresponding to the scaling in eq.~\eqref{eq:moff_pred}. 
After a short initial growth---the larger $N$, the shorter this transient stage---, 
the nonlinear simulations exhibit a decay of $b_{33}$. This indicates that the vertical kinetic energy 
decays faster than the total kinetic energy, in agreement with figure \ref{fig:uvw}. 
This observation alone could lead to the incorrect conclusion that the quadratic nonlinearity 
in eq.\eqref{eq:momentum} tends to restore 3D isotropy. 
A closer inspection of $b_{33}$ provides a significantly different viewpoint, when considering 
the decomposition of $b_{33}=b_{33}^e+b_{33}^{\mathcal{Z}}$ in directional and polarization anisotropy 
contributions \cite{camb89,CAMBON-MANSOUR-GODEFERD}, following:
\begin{align}
\label{eq:b33e}
b_{33}^e & =\frac{1}{2\mathcal{K}}\int\left(e(\bm{k})-\frac{E(k)}{4\pi k^2}\right)\sin^2\theta\textrm{d}^3\bm{k} \\
\label{eq:b33z}
b_{33}^{\mathcal{Z}} & =\frac{1}{2\mathcal{K}}\int\mathcal{Z}(\bm{k})\sin^2\theta\textrm{d}^3\bm{k}
\end{align}
where $\theta$ is the polar angle between the wave vector $\bm{k}$ and the axis of symmetry, $E(k)$ is the spherically-averaged kinetic energy spectrum, $\mathcal{Z}(k)$ is the polarization 
spectrum (see below).
This decomposition is easier described in polar-spherical coordinates (see figure \ref{fig:scheme_ezh}) with
unit toroidal vector $\bm{e}^{(1)}$ and poloidal vector $\bm{e}^{(2)}$. 
Due to incompressibility $\hat{\bm{u}}(\bm{k})\cdot\bm{k}=0$, $\hat{\bm{u}}=\hat{u}^{(1)}\bm{e}^{(1)}+\hat{u}^{(2)}\bm{e}^{(2)}$: each Fourier mode 
has a toroidal contribution $\hat{u}^{(1)}$ and a poloidal one $\hat{u}^{(2)}$ (we drop the explicit
$\bm{k}$ dependence in this paragraph). 
Assuming horizontal plane mirror symmetry, thus without helicity, the kinetic energy density is 
\begin{equation}
e=E^{\textrm{pol}}+E^{\textrm{tor}}=\frac{1}{2}\left(\hat{u}^{(2)*}\hat{u}^{(2)}+\hat{u}^{(1)*}\hat{u}^{(1)}\right)
\end{equation}
 and the polarization tensor density is
\begin{equation}
\mathcal{Z}=E^{\textrm{pol}}-E^{\textrm{tor}} \ .
\end{equation} 
A detailed presentation of this decomposition can be found in \cite{camb89}. 
\textcolor{black}{Note that $b_{33}^e$ is close to the so-called Shebalin angle \cite{sheb83},
whereas $b^e_{ij}$ is shown \cite{CAMBON-MANSOUR-GODEFERD} to be exactly minus half the non-dimensional
deviator of the ``dimensionality tensor" \cite{kass99}.}

Figure \ref{fig:b33}(b) shows the time evolution of $b_{33}^e$ and $b_{33}^{\mathcal{Z}}$. 
For all values of $N$, $b_{33}^e$ increases indicating that the kinetic energy density 
$e(\bm{k})$ is not isotropically distributed among the different wavevector orientations. 
In the linear runs (symbols in figure \ref{fig:b33}) $b_{33}^e$  follows closely the nonlinear evolution
 (continuous lines in figure \ref{fig:b33}). Unlike $b_{33}^e$, a departure is observed for $b_{33}^{\mathcal{Z}}$: 
it decreases in the nonlinear simulations, but remains negligible in the linear runs. 
Thus, the decay of $b_{33}$ observed in figure \ref{fig:b33}(a) is due to a nonlinear decay 
of the polarization contribution $b_{33}^{\mathcal{Z}}$.
\begin{figure}[t!]
\includegraphics[width=\columnwidth]{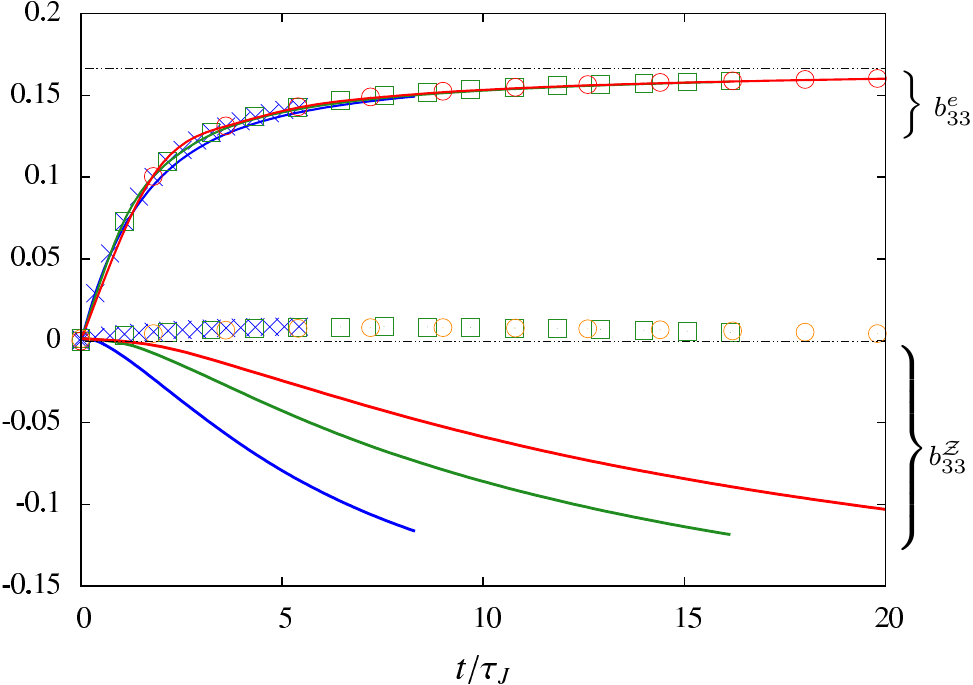}
\caption{Same as figure \ref{fig:b33}(b) but with time normalized by the Joule timescale $\tau_J=\eta/B_0^2$.}
\label{fig:b33adim}
\end{figure}

The analysis can be even more refined when considering the scale dependence
of the poloidal/toroidal anisotropy.
In eq.\eqref{eq:momentum}, one observes that the Lorentz force 
introduces an additional anisotropic dissipation, upon examining the
 Fourier transform of $\bm{F}$:  
$\hat{\bm{F}}=-\left(\bm{B}_0\cdot\bm{k}\right)^2\hat{\bm{u}}/\left(\eta k^2\right)$ shows that 
the vertical modes ($\bm{k}//\bm{B}_0$) are strongly damped whereas horizontal 
ones ($\bm{k}\perp\bm{B}_0$) are only modified by pressure effects. 
This well-known phenomenon, which is linear and characterized by a Joule timescale $\tau_J=\eta/B_0^2$, is responsible for the initial growth of $b_{33}^e$ observed in figure \ref{fig:b33}(a). \textcolor{black}{Note that $\tau_J$ can be used to rescale $t$ (as in figure \ref{fig:b33adim}) so that all curves for $b_{33}^e$ collapse. However, with this scaling, the evolution of $b_{33}^{\mathcal{Z}}$ is still very different for the various cases. As $N$ increases (\textit{i.e.} the nonlinear interactions become more and more negligible compared to linear ohmic dissipation), the decay of $b_{33}^{\mathcal{Z}}$ is slower, clearly indicating that the polarization effect is triggered by a nonlinear phenomenon.}

\subsection{Angular energy spectra}

\begin{figure*}
\includegraphics[width=\columnwidth]{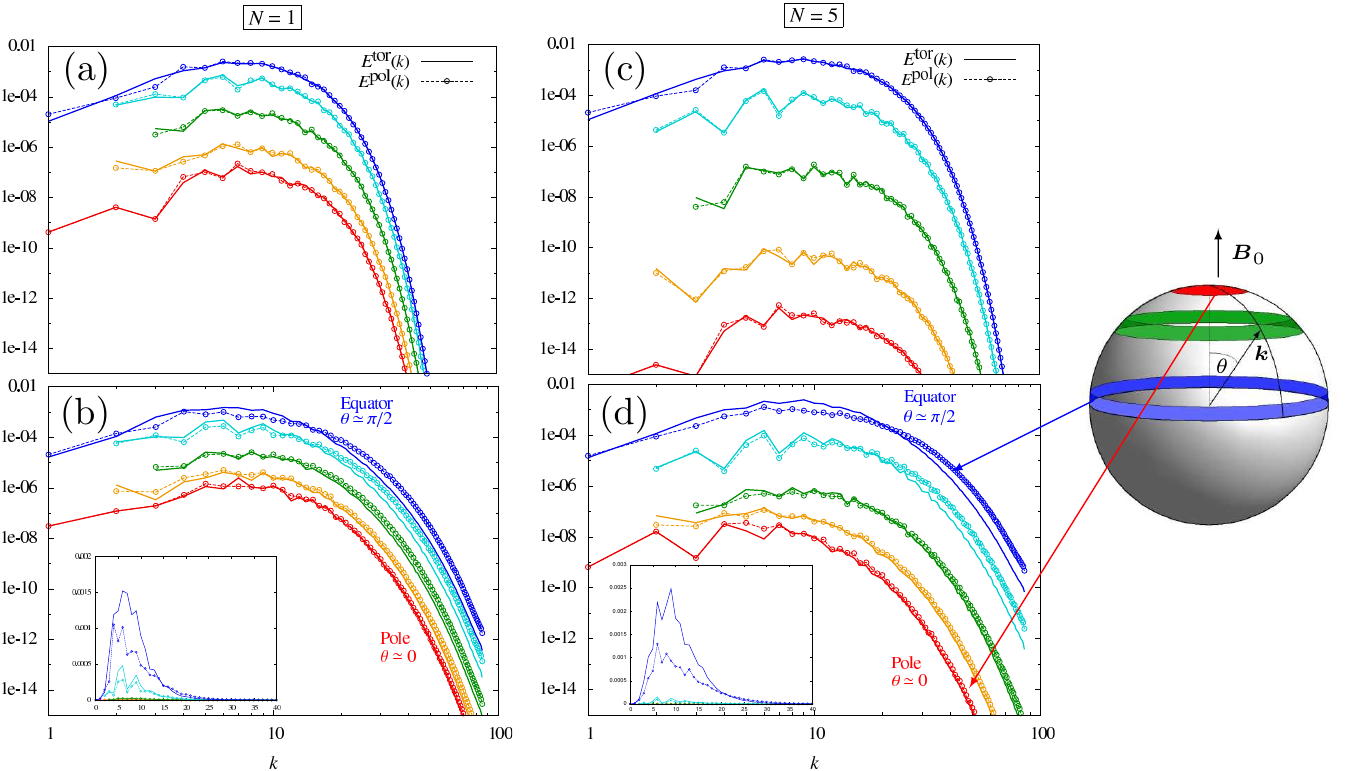}
\caption{\textcolor{black}{Angular energy spectra (note that for the case $N=5$, the results are plotted for $t\simeq20\tau_J$ whereas for the case $N=1$, it corresponds to $t\simeq5\tau_J$). Top: Linear simulations. Bottom: Non-linear simulations. Left (figures (a) and (b)): $N=1$. Right (figures (c) and (d)) $N=5$. On figures (b) and (d), the embedded figures corresponds to the same results presented in linear scale.}}
\label{fig:spang_lrm}
\end{figure*}

Let us define angular energy spectra defined by
\vspace{-1mm}
\begin{equation}
\label{eq:ekt_def2}
E(k,\theta)=f(\theta)\sum_{T(k,\theta)} \hat{\bm{u}}\cdot\hat{\bm{u}}^* \ ,
\vspace{-2mm}
\end{equation}
where $T(k,\theta)$ denotes the torus-shaped volume defined from $k-\Delta k/2<|\bm{k}|<k+\Delta k/2$ and 
$\theta-\Delta\theta/2<\theta<\theta+\Delta\theta/2$,
with $\Delta k=1$ and $\Delta\theta=\pi/10$ the Fourier space discretization steps 
(blue domain in figure \ref{fig:scheme_ezh}). $f(\theta)$ is the geometrical weighting function such that the angular spectra collapse 
in isotropic turbulence. Torus-averaged angular spectra $E(k,\theta)$ have already been used in the context 
of rotating turbulence \cite{CAMBON-MANSOUR-GODEFERD} and are similar to the so-called 
ring decomposition \cite{bura08}. On figures \ref{fig:spang_lrm}, we present the angular spectra for five orientations
 from the equator ($\theta\simeq\pi/2$) to the pole ($\theta\simeq0$), for $N=1$ (figures \ref{fig:spang_lrm}(a) and (b)) and $N=5$ (figures \ref{fig:spang_lrm}(c) and (d)). 
In addition, we distinguish the poloidal and toroidal spectra $E^{\textrm{pol}}$ and  $E^{\textrm{tor}}$, and the linear simulations from the nonlinear ones (figures \ref{fig:spang_lrm}(a)(c) and figures \ref{fig:spang_lrm}(b)(d) respectively). The figures show that almost all the energy is concentrated in the equatorial spectrum, as a result from the linear Joule dissipation, independently on the poloidal or toroidal contributions, 
and in both the linear and nonlinear simulations.
The reasons for the observed differences between the latter two cases are two-fold:  
\textit{(a)} the nonlinear downscale energy cascade increases the energy in the small scale
range of the nonlinear runs with respect to the linear ones; 
\textit{(b)} in the meantime, a nonlinear angular transfer kicks in, explaining the polar energy depletion, in the same figures, for the
linear runs (see \textit{e.g.} \cite{bura08}).
Given that polar modes contribute only to the horizontal kinetic energy 
(clearly from geometrical reasons,
see fig.\ref{fig:scheme_ezh}) combined with a more efficient Joule dissipation in the same region, 
this linear mechanism is responsible for the rapid decrease of horizontal energy, 
\textit{i.e.} rapid \textit{increase} of $b_{33}$,  observed at short times on fig.\ref{fig:b33}(a). 
There is however nothing in this mechanism that explains the decay, of $b_{33}^{\mathcal{Z}}$ in the nonlinear simulation results plotted on fig.\ref{fig:b33}(b).

The crucial difference between linear and nonlinear simulations in figures \ref{fig:spang_lrm} is brought to light
by the poloidal-toroidal decomposition. At the equator in the full DNS, 
the horizontal kinetic energy (also $E^{\textrm{tor}}$ at this specific 
orientation) 
is dominant at large scales, whereas the vertical kinetic energy 
(also $E^{\textrm{pol}}$) is dominant at small scales 
\textcolor{black}{(the small plot in inset of figures \ref{fig:spang_lrm}(b) and (d), in linear scale, shows more clearly the important 
large-scale energy
gap between the two modes)}.
This result explains the increase of horizontal kinetic energy with respect to the vertical energy, 
hence the decay of $b_{33}^{\mathcal{Z}}$. It is clearly of nonlinear origin since it disappears 
when nonlinear interactions are removed (see figures \ref{fig:spang_lrm}(a) and (c)). 
Although already observed in previous works \cite{bura08_2}, this unpredicted 
dominance of horizontal energy was not understood. We provide here a more precise analysis 
since only modes such that $\bm{k}\!\perp\!\bm{B}_0$ appear to have noticeably different poloidal and
toroidal energies.
\begin{figure}
\includegraphics[width=\columnwidth]{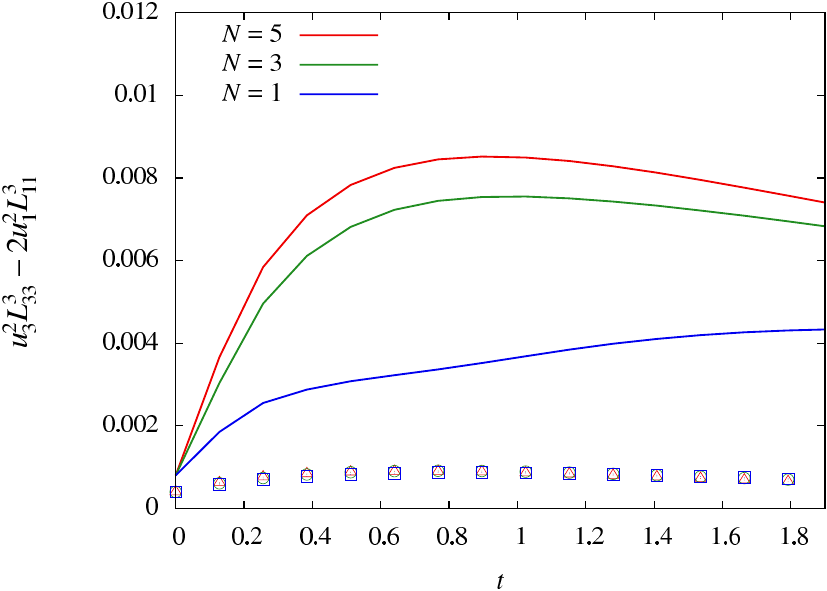}
\caption{Evolution with time of the quantity $u_3^2L_{33}^3-2u_1^2L_{11}^3$. The nonlinear results are represented in continuous lines and linear results in symbols.}
\label{fig:lij}
\end{figure}

\subsection{Correlation lengths}
\label{sec:corr_l}

\textcolor{black}{
We present in this paragraph some results about correlation lengths, which can be computed directly from the angular dependent spectra,
and we thus check that our simulations are free from any artificial confinement due to periodic boundary conditions.
Let us introduce the velocity correlation lengths defined by
\begin{equation}
L_{ij}^l=\frac{1}{\left<u_iu_j\right>}\int_0^{\infty}\left<u_i(\bm{x})u_j(\bm{x}+\bm{r})\right>\textrm{d}r
\end{equation}
where $r_k=r\delta_{kl}$ is the two-point separation. In the current axisymmetric configuration, the most relevant anisotropy indicators involve the integral length scale with vertical separation but relative to either vertical or horizontal velocity components \cite{camb89}:
\begin{align}
\label{eq:defl333}
L_{33}^3 & =\frac{2\pi^2}{u_3^2}\int_0^{\infty}\left[e(\bm{k})+\Re\mathcal{Z}(\bm{k})\right]\Big|_{k_z=0}k\textrm{d}k \\
\label{eq:defl113}
L_{11}^3 & =\frac{\pi^2}{u_1^2}\int_0^{\infty}\left[e(\bm{k})-\Re\mathcal{Z}(\bm{k})\right]\Big|_{k_z=0}k\textrm{d}k \ .
\end{align}
The correlation length $L_{11}^3$ is characterized by the most critical growth rate, as qualitatively observed on visualizations \ref{fig:visus}(b) and (d). At the end of the simulations (\textit{i.e.} for $t\approx1.8$), $L_{11}^3\approx2, \ 1.6, \ 1$ for the respective cases $N=5, \ 3, \ 1$. In all cases, $L_{11}^3$ is significantly smaller than the box size $2\pi$, except for $N=5$ for which the largest integral scale is about one third of the numerical box size. However, we have seen that the results for the case $N=1$ and $N=5$ are qualitatively very similar (see for example figures \ref{fig:spang_lrm}). We are therefore confident in the fact that our simulations are free from any spurious periodic effects.} 

\textcolor{black}{
Moreover, it is possible to isolate the contribution due to polarization looking at the quantity:
\begin{equation}
u_3^2L_{33}^3-2u_1^2L_{11}^3=\int_0^{\infty}4\pi^2\Re\mathcal{Z}(\bm{k})\Big|_{k_z=0}k\textrm{d}k \ ,
\end{equation}
plotted on figure \ref{fig:lij}. This quantity has two main advantages: computing from equations \eqref{eq:defl333} and \eqref{eq:defl113}, its departure from zero is only due to the polarization $\mathcal{Z}(\bm{k})$; secondly, this quantity is accessible experimentally and is thus of particular interest. Again, one observes on figure \ref{fig:lij} that polarization is negligible in the linear case whereas a significant growth is observed in the nonlinear case.}

In short, the anisotropic Lorentz force is responsible for a 
preferential dissipation leading to a turbulent flow independent of the vertical direction (see figure \ref{fig:visus}(b)); this linear effect concentrates the kinetic energy among horizontal modes, as attested by the growth of $b_{33}^e$. Then, due to nonlinearity, the energy is anisotropically distributed among poloidal and toroidal components leading to a nonzero polarization and negative $b_{33}^{\mathcal{Z}}$.

\section{The analogy with 2D-3C turbulence}
\label{sec:2d3c}

The quasi-static MHD turbulence is however far from being  purely two-dimensional. Although the flow tends to be invariant in the vertical direction, 
\textcolor{black}{the vertical kinetic energy is not negligible compared to the horizontal one} (and is even dominant in the linear regime). 
This state is often referred to as two-dimensional, three-components (2D-3C) or ``two-and-a-half-dimensional" turbulence \cite{mont82}. 
We now propose a simple description of the above results on directional anisotropy and polarization.
We first write the Navier-Stokes equations for a flow independent of the vertical direction $z$, 
so that $\partial/\partial z=0$:
\begin{align}
\frac{\partial w}{\partial t}+\bm{u}_{\perp}\cdot\nabla_{\perp}w & =\nu\nabla_{\perp}^2w \label{eq:nsuz}\\
\frac{\partial \bm{u}_{\perp}}{\partial t}+\bm{u}_{\perp}\cdot\nabla_{\perp}\bm{u}_{\perp} & =-\frac{1}{\rho}\nabla_{\perp} p+\nu\nabla_{\perp}^2\bm{u}_{\perp} \label{eq:nsw}
\end{align}
where  $\bm{u}_{\perp}=\left(u,v\right)$ is the horizontal velocity component and $\nabla_{\perp}$ 
 the horizontal gradient. In both equations, the Lorentz force disappears altogether. 
Eq.(\ref{eq:nsuz}) for the vertical velocity is that of a passive scalar,
 whereas eq.(\ref{eq:nsw}) for  horizontal velocity is characteristic of 2D \textcolor{black}{hydrodynamic} turbulence.
One therefore expects an inverse energy cascade  for the horizontal velocity (or at least a strongly attenuated direct cascade) and a classical direct cascade 
for the vertical velocity. This is consistent with the results observed on figures \ref{fig:spang_lrm}(b) and (d) 
since the slope of the toroidal (\textit{i.e.} horizontal) component is steeper than 
the slope of the poloidal (\textit{i.e.} vertical) component for horizontal modes 
(\textit{i.e.} at the equator). Thus, we explain the large-scale dominance of horizontal energy 
as a result of 2D-like cascade for the horizontal velocity.
The vertical component behaves like a passive scalar, and is therefore 
characterized by a direct cascade so that the vertical energy is dominant at small scales. 
Note that this effect has nothing to do with a restoration of isotropy, 
even if $b_{33}$ tends to its isotropic value.

\begin{figure}
\includegraphics[width=\columnwidth]{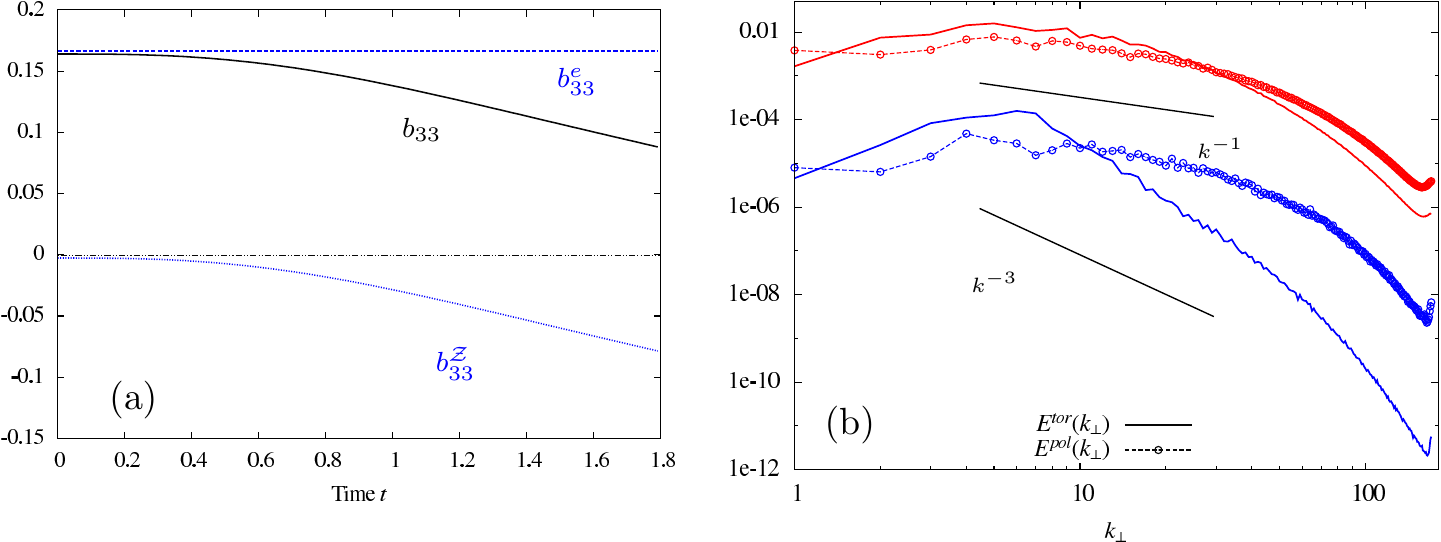}
\caption{\label{fig:2D-3C} (a) $b_{33}(t)$ from a 2D-3C simulation. (b) Equatorial energy spectra $E^{\textrm{pol,tor}}(k_{\perp})$. Red/top: Quasi-static 3D MHD using $512^3$ Fourier modes. Blue/bottom: 2D-3C simulation using $512^2$ Fourier modes (shifted down by two decades).}
\end{figure}

To assess the validity of this analysis, we perform two additional simulations.
First, we compute the evolution of hydrodynamic turbulence from an initial 2D-3C state and a resolution of $512^2$. Figure \ref{fig:2D-3C}(a) shows that $b_{33}^e=1/6$ during all the simulation. As in quasi-static MHD turbulence, 2D-3C turbulence is characterized by a negative polarization indicating a dominance of toroidal kinetic energy with respect to the poloidal one. Figure \ref{fig:2D-3C}(b) shows the equatorial toroidal and poloidal energy spectra $E^{\textrm{tor,pol}}(k_{\perp})$ (bottom of the figure) that again exhibit a dominance of toroidal energy 
at large scales and the opposite at small scales, in support to our previous analysis 
on the final state of quasi-static MHD turbulence.
Secondly, we would like to validate the universality of the mechanism at higher
Reynolds number MHD turbulence, by performing a quasi-static simulation at $\textit{Re}\simeq300$,
using $512^3$ Fourier modes. We choose the case $N=5$ and $R_M=0.1$. In this 3D simulation, the crossing of the toroidal and poloidal spectra (top spectra in figure \ref{fig:2D-3C}(b)) appears as in figures \ref{fig:spang_lrm}(b) and (d) and in the previously described 2D-3C spectra. The $k^{-3}$ and $k^{-1}$ slopes are plotted for comparison
with common scalings of 2D turbulence with passive scalar \cite{batch59,bos09}, again showing 
clearly that the energy cascade is more efficient for the toroidal energy than for the poloidal one, resulting in a steeper slope for toroidal kinetic energy. \textcolor{black}{Of course, many differences arise looking at figure \ref{fig:2D-3C}(b). This is due to the fact that quasi-static turbulence is not purely invariant in the vertical direction (see visualization on fig.\ref{fig:visus}(b)).}

\section{Conclusion}

\textcolor{black}{In conclusion, we have found indications that the final state of quasi-static MHD turbulence, reached after a time sufficient for
developping a strong anisotropy, is analogous to 2D-3C hydrodynamic turbulence.}
The mechanism responsible for the
transition from 3D isotropic turbulence 
to 2D-3C turbulence is subtle. It combines the anisotropic Joule dissipation, which is a linear  
phenomenon, and nonlinear energy transfers. If nonlinearities are neglected,  
this final state is characterized by a dominance of vertical kinetic energy \cite{moff67}.  
However, from the quasi-2D state arising from Joule dissipation, 
nonlinearity induces different dynamics for toroidal/horizontal velocity components 
and for poloidal/vertical components. 
The horizontal flow behaves like 2D turbulence whereas the vertical flow behaves like a passive scalar \textcolor{black}{advected by 2D turbulence}. 
Accordingly, the energy cascade (and thus the dissipation) is more efficient in the vertical direction, explaining the overall dominance
 of horizontal energy. We have also shown that the poloidal/toroidal decomposition 
and the distinction between directional and polarization anisotropy are fundamental to explain
these physical phenomena. This description may help to explain the large scale anisotropy
of some geophysical or astrophysical conducting fluid flows, with the possibility of
coupling the 2D-3C model for the large scales to a specific turbulence closure at small scales,
thus allowing to achieve very high Reynolds number simulations.

The authors thank the computing center IDRIS of CNRS for the allocation of cpu time under project numbers 071433 and 022206.
%
%
%
%

%
%
\end{document}